    \documentclass[12pt,psf,epsf]{article}
\usepackage[centertags]{amsmath}
\usepackage{amssymb}
\usepackage{graphicx}
\usepackage{epsfig}
\usepackage{ulem}
\usepackage[english]{babel}
\usepackage{array}
\usepackage{amsthm}
\usepackage{latexsym}
\usepackage[mathcal]{euscript}
\pdfoutput=1
\usepackage{epsfig}

 \usepackage{jheppub}
 \usepackage{hyperref}

\usepackage{subfigure}
\usepackage{psfrag}

\usepackage{epsfig}
\usepackage[latin1]{inputenc}
\usepackage{float}
\usepackage{graphicx}
\usepackage{cancel}
\usepackage{mathrsfs}
\usepackage{amssymb}
\usepackage{amsfonts}
\usepackage{amsmath}
\usepackage{slashed}
\usepackage{graphicx}
\usepackage{bm}
\usepackage{color}
\newcommand{\be}{\begin{equation}}
\newcommand{\ee}{\end{equation}}
\newcommand{\bea}{\begin{eqnarray}}
\newcommand{\eea}{\end{eqnarray}}
\newcommand{\bwt}{\begin{widetext}}
\newcommand{\ewt}{\end{widetext}}

\newcommand{\nn}{\nonumber}

\newcommand{\bi}{\begin{itemize}}
\newcommand{\ei}{\end{itemize}}

\newcommand\cC{{\cal C}}
\newcommand\cV{{\cal V}}
\newcommand\cL{{\cal L}}

\usepackage{setspace}

\begin{document}
%\doublespacing

\title {On the  entanglement  contour of excited states in the holographic CFT}

\author{Dmitry S. Ageev}
\affiliation{Department of Mathematical Methods for Quantum Technologies, Steklov Mathematical Institute of Russian Academy of Sciences, Gubkin str. 8, 119991 Moscow, Russia }

\emailAdd{ageev@mi-ras.ru}

\abstract{In this paper, we study the entanglement contour in a general excited state in the holographic 2d CFT using the partial entanglement entropy proposal. We show how  the thermodynamics fixes the entanglement contour relating it to  first law of entanglement. We derive the entanglement contour for a general time-dependent excited state and consider a quenched initial state in the presence of
spatial boundaries as an explicit example. Finally, we  comment on the coarse-graining and the complexity contour in the $AdS_3/CFT_2$.  }

\maketitle

\newpage

\section{Introduction}
The connection between quantum information and quantum gravity in the context of the AdS/CFT correspondence has played a growing role in our understanding of these phenomena\cite{Ryu:2006bv}-\cite{Swingle:2009bg}. In the AdS/CFT correspondence  the  (Hubeny--Rangamani--Ryu--Takayanagi) HRRT formula establish the connection between spacetime geometry and entanglement entropy \cite{Ryu:2006bv},\cite{Hubeny:2007xt}. This formula relates the entanglement entropy of the fixed spatial subregion $A$  and the  extremal codimension-2 hypersurface with  area ${\cal E}_A$ spanned on the boundary of this subregion
\be \label{RT}
S(A)=\frac{1}{4G}{\cal E}_A.
\ee 
The entanglement entropy is an important quantity that captures non-local correlations associated with the degrees of freedom corresponding to  $A$. There is a natural question about decomposition of such  non-local observable into less coarse-grained quantity. The step towards  such refinement has been made in \cite{chen-vidal} where the definition of entanglement contour was given and studied further in \cite{Coser:2017dtb}-\cite{Han:2019scu}.  The entanglement contour $f_A(x)$ in its essence  is the function associated to the fixed subsystem $A$ which being integrated over $x \in A$ gives the entanglement for $A$
\be 
S=\int_{x \in  A} f_A(x)dx.
\ee 
Also, there are different restrictions like positivity and normalization that $f_A(x)$ has to satisfy. The total list of these restrictions is unknown at the moment and the entanglement contour is fixed by the existing list in a non-unique way. 

Recently the entanglement contour in holographic description based on the partial entanglement entropy was proposed  in \cite{Wen:2018whg} and considered further in \cite{Wen:2019ubu},\cite{Kudler-Flam:2019nhr}. In \cite{Kudler-Flam:2019nhr}  the contour function in some lattice models has been investigated and the results are consistent with the partial entanglement proposal.
Here we study different aspects of this proposal. We give the description of the entanglement contour for a general excited state dual to the Banados geometry \cite{Banados:1998gg}. In this case the entanglement contour is fixed by the stress-energy tensor in dual theory. It is known that there is a general relation \cite{Bhattacharya:2013bna,Nozaki:2013vta} between thermodynamics and the entanglement entropy of small subsystems. We find that the similar "entanglement contour law" thermodynamic description of  the subsystem also takes place and write it down explicitly using the results of \cite{Bhattacharya:2013bna,Nozaki:2013vta}.

From the definition and properties of the entanglement contour, it is natural to think that the holographic definition is related to some geometric quantities defining the entanglement.  In this paper we focus on the $AdS_3$ for simplicity. In this case, this leads us to the construction of the measure from the HRRT length element similar to the entanglement contour. The straightforward integration of this measure does not give us the entanglement entropy. However, after the proper choice of the cutoff associated with the particular state  and subsystem, we get the correct answer. The distribution of entanglement according to this measure reproduces all qualitative features of the entanglement contour derived using (for example) partial entanglement proposal. We extend construction to the non-equilibrium state and call this measure the geometric entanglement contour. We calculate it for the CFT dual of the Banados geometry explicitly. Note, that more fine-grained correspondence  \cite{Wen:2018whg,Wen:2019ubu} relating points of the interval and on the geodesic is known following partial entanglement entropy proposal.

Besides the entanglement entropy, different quantum non-local measures are known.   Recently it has been proposed to describe the evolution of hidden degrees of freedom in holographic systems using a notion of quantum complexity. There are different proposals on the definition of this quantity in holography and quantum systems \cite{Stanford:2014jda}-\cite{Caputa:2017urj}. At the moment, there is no conventional description of complexity in quantum field theory as well as in holography.  In this paper, we discuss the analog of entanglement contour for CV complexity. We call this quantity the complexity contour focusing on the peculiar example of the $AdS_3/CFT_2$ case. It is known that the dependence of the CV complexity on the interval length is in its essence temperature-independent  \cite{Abt:2017pmf}. The construction similar to the partial entanglement does not clarify the situation leading to temperature-independent answer. However, one can proceed with the geometrical definition that leads to the non-trivial "complexity degrees of freedom" distribution. This shows that the geometric definition of contour-like measures is an important thing to study on its own.

This paper is organized as follows. In section \ref{entc} we review the entanglement contour proposal and consider different simple examples of it. In section \ref{entE1} we obtain the entanglement contour for the arbitrary excited state dual to the Banados geometry. In section \ref{entE2} we consider the thermodynamic description of the entanglement contour in the spirit of first law of entanglement. We introduce the geometric contours based on geodesics in section \ref{GCM}. We finalize with the comments on the importance of the
geometric definition of the contour functions in holography and discuss complexity
contour proposals in section \ref{CM}.

\section{The entanglement contour and geometric entanglement contour: definitions and basic examples}\label{entc}
As it was mentioned in the introduction, the entanglement contour associated with the subsystem  $A$ is the function $f_A(x)$  defined as
\be \label{EC}
S(A)=\int_{x \in  A} f_A(x)dx,
\ee 
where $S(A)$ is the entanglement entropy of $A$. For simplicity, let us restrict our attention to the case where $A$ is the connected region (for example, single interval or disc, etc.). Apart normalization condition \eqref{EC} the function $f_A(x)$ has to satisfy some properties to be the entanglement contour: 
\begin{itemize}
    \item Entanglement contour is a positive function: 
\be     
    f_A(x)>0,
\ee    
   
    \item The entanglement contour $f_A(x)$ should inherit spatial symmetries of  the reduced density matrix $\rho_A$.
    \item Invariance under local unitary transformations.
    \item Upper bound: if $ {\cal H}_{T} = {\cal H}_{B} \otimes {\cal H}_{\bar B}$ and ${\cal H}_X \subseteq {\cal H}_{B} $ then
    \begin{gather}
        f_{T}(x) \leq S(B).
    \end{gather}
    % Here, $S(\Omega_A)$ is the entanglement entropy corresponding to the subspace $\mathcal{H}_{\Omega_A}$.
\end{itemize}
Recently in  \cite{Wen:2018whg} the proposal for the contour function based on the notion of  the partial entanglement entropy has been given.  Let us consider some one-dimensional system and the subsystem $A$. The partial entanglement  entropy  $s_A(A_2)$ is  defined as 
\begin{gather} \label{partS}
A=A_1\cup A_2\cup A_3,\\\nn
s_A(A_2)=\frac{1}{2}\Big(S(A_1 \cup A_2)+S(A_2 \cup A_3)-S(A_1)-S(A_3 )\Big).
\end{gather} 
Roughly speaking  partial entanglement entropy gives us the contribution of $A_2$ to the entanglement of the total system $A$. The generalization of \eqref{partS} to the case of n-subpartitions $A_i$  is given \cite{Kudler-Flam:2019oru} by
\begin{gather}  
\underset{i}{\cup}A_i=A,\\ \nn
s_A(A_i)=\frac{1}{2}\Big(S(A_i|A_1\cup...A_{i-1})+S(A_i|A_{i+1}\cup...A_{n})\Big),
\end{gather}
where $S(A|B)$ is the conditional entropy defined as
\be \label{partial}
S(A|B)=S(A\cup B)-S(B).
\ee 
To get the entanglement contour $f_A(x)$ for the ground state of 2d CFT and  when $A$ is a single interval of the length $\ell$ such that $x \in (-\ell/2,\ell/2)$ we proceed as follows. We consider $x_1>-\ell/2$ and $x_2=x_1+\delta x$, such that $A_1\in(-\ell/2,x_1)$, $A_2 \in (x_1, x_2)$ and $A_3 \in (x_2,\ell/2)$.
Using formula \eqref{partial} and expanding it to the leading order in $\delta x$ we get
\be 
S_A(x_1)\approx f_A(x_1)\delta x,\,\,\,\,\,\,f_A(x_1)=\frac{c}{6} \frac{4\ell}{\ell^2-4x_1^2},
\ee 
where  $f_A(x_1)$ is the entanglement contour  ground state.
The entanglement entropy for 2d CFT at finite temperature of the interval of length $\ell$ is given by
\be 
S(\ell)=\frac{c}{3}\log \left(\frac{\beta}{\pi \epsilon}\sinh\frac{\beta}{\pi}\ell\right),
\ee 
leading us to the entanglement contour for the thermal state 
\be \label{c-thermal}\nn
f_A(x)=\frac{c}{3}\frac{\pi}{\beta}\left(\coth\left(\frac{\pi}{\beta}\left(\frac{\ell}{2}-x\right)\right)+\coth\left(\frac{\pi}{\beta}\left(\frac{\ell}{2}+x\right)\right)\right).
\ee
These examples have been considered in \cite{Wen:2018whg,Kudler-Flam:2019oru}.
Formula for the CFT defined on the circle of length $L$ is given by \eqref{c-thermal} up to the change  $\beta\rightarrow L$ and $\coth \rightarrow \cot$.

\section{The entanglement contour for excited states}\label{entE}
\subsection{The excited states dual to Banados metric}\label{entE1}
\begin{figure}[h!]
\centering
\includegraphics[width=6.5cm]{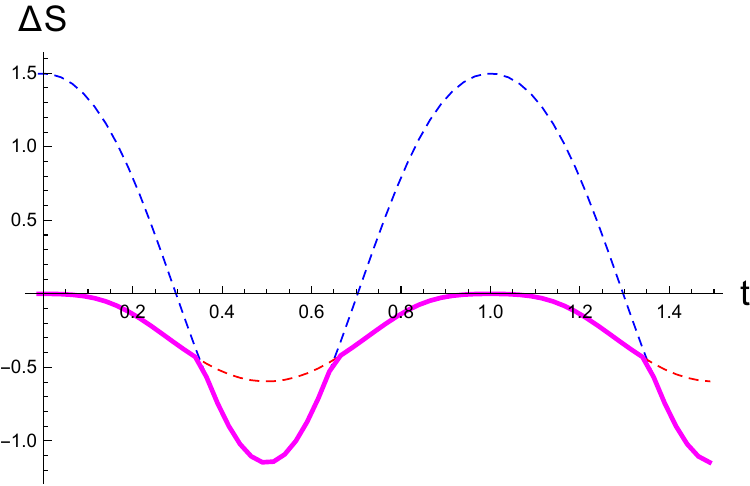}
 \caption{Evolution of the difference between the entanglement entropy at finite volume and temperature corresponding to $\beta=1, L=1$ and zero-temperature entanglement with $L=1$. Blue and red curves correspond to different competing geodesics (channels on the CFT side). The magenta curve corresponds to the resulting entanglement entropy. }
 \label{fig:ds}
\end{figure}

\begin{figure}[h!]
\centering
\includegraphics[width=5.5cm]{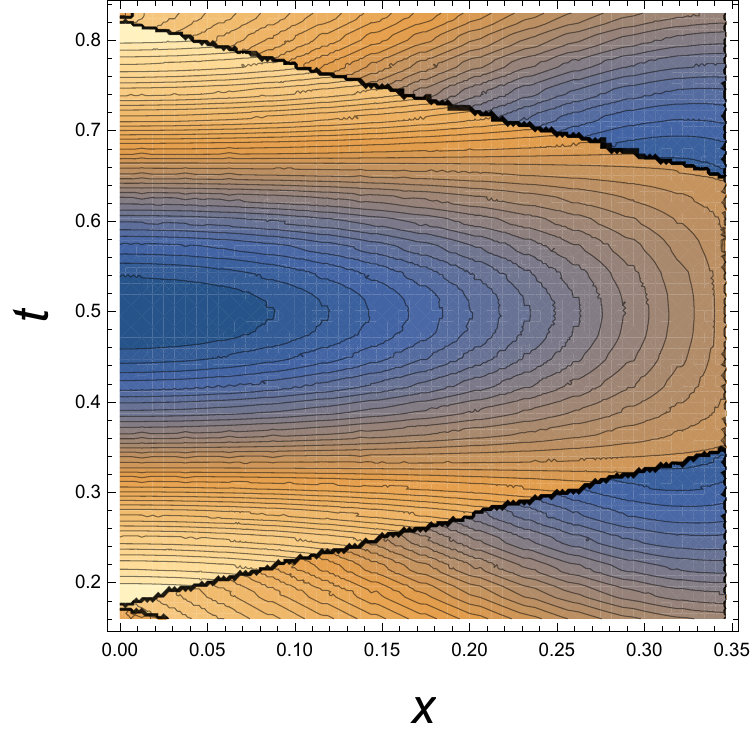}
 \caption{Evolution of the difference between the entanglement contour of $\beta=1, L=1$ and zero-temperature  contour with $L=1$. The size of the interval is fixed at $\ell\in(-0.35,0.35)$. Here both connected and disconnected geodesic channels are included. }
 \label{fig:osc1}
\end{figure}

\begin{figure}[h!]
\centering
\includegraphics[width=4cm]{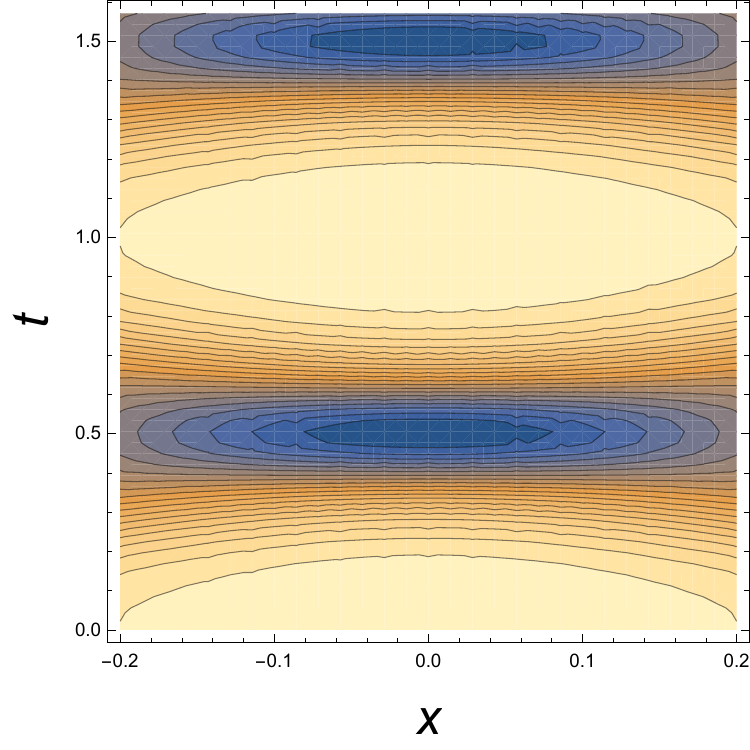}
\includegraphics[width=4cm]{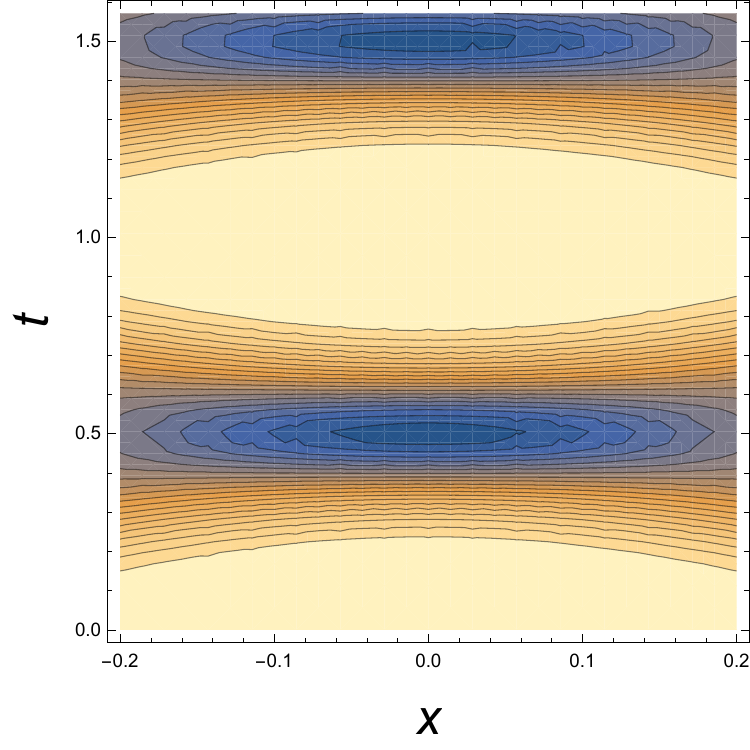}
 \caption{Left plot: the difference between  the entanglement contour for the length $\ell$ interval defined by \eqref{finite}  and the entanglement contour of the ground state 2d CFT. Right plot: the same for geometric entanglement contour.  Here $L=1$, $\beta=0.8$ and $\ell=0.4$. }
 \label{fig:osc2}
\end{figure}

In the previous  section we have considered the basic examples of the stationary metrics dual to different simple states of the 2d CFT. Now we turn to the Banados metric describing quite general class of the CFT states (see \cite{Sheikh-Jabbari:2016unm,Sheikh-Jabbari:2016znt} for discussion of the Banados geometry and entanglement structure of its holographic dual). This metric has the form
\begin{gather}
     \label{banados}
ds^2=L^2\Big(\frac{dz^2}{z^2}+\Big(\frac{1}{z^2}+\frac{z^2}{16} {\cal L}_-(x_-) \cL_+(x_+)\Big)dx_-dx_++\frac{\cL_-(x_-)}{4}dx_-^2+\frac{\cL_+(x_+)}{4}dx_+^2\Big),
\end{gather} 
where $\cL_-$ and $\cL_+$ are the arbitrary functions of the single coordinate $x_-$ and $x_+$ respectively. These functions are proportional to the CFT stress-energy tensor one-point function
\be 
\langle T_{\pm\pm} \rangle=\frac{L}{16 G} \cL_{\pm }.
\ee
 and one of the main advantages of the Banados metric is that one can obtain it from the Poincare metric
\begin{gather}
ds^2=\frac{1}{u^2}\left(-dy_-dy_++du^2\right),\,\,\,\,\,\,\,\,y_\pm=y\pm t,
\end{gather} 
using mappings
\begin{gather} \label{map1}
y_\pm = F_{\pm}\left(x_\pm\right)-\frac{2 z^2 F_{\pm}'{}^2
   F_{\pm}''}{z^2 F_{\mp}''\left(x_-\right) F_{\pm}''+4
   F_{\mp}' F_{\pm}'},\,\,\,\,\,\,\,
u=\frac{4 z \left(F_-' F_+'\right){}^{3/2}}{z^2
   F_+'' F_-''+4 F_-'
   F_+'},
\end{gather}
which act near the boundary $u,z\rightarrow 0$ as 
\begin{gather}
 \label{nbmap1}
y_\pm = F_\pm\left(x_\pm\right),\,\,\,\,\,\,
u=z\sqrt{f_-(x_-)f_+(x_+)},\,\,\,\,\,\,u,z\rightarrow 0.
\end{gather}
 Factors ${\cal L}_\pm$ defining stress-tensor are related to the mapping $F$ by the Schwarzian  derivative $S$
 \be 
{\cal L}_\pm=-S(F_{\pm}(x_\pm)),\,\,\,\,S(f(x))=\frac{f^{'''}(x)}{f^{'}(x)}-\frac{3}{2}\left(\frac{f^{''}(x)}{f^{'}(x)}\right)^2.
 \ee 
  In fact to get the metric with the boundary stress-tensor defined by the Schwarzian of $F_\pm$ it is enough to use the mappings \eqref{nbmap1} 
\begin{gather}
y_\pm=f(x_\pm),\,\,\,\,\,\,
u=z\sqrt{f_-(x_-)f_+(x_+)},
\end{gather}
as it was noted in \cite{Mandal:2014wfa}. This shows large gauge freedom of asymptotically  $AdS_3$ space. For  simplicity let us restrict our attention to the case when $F_+=F_-=F$.
In principle this mapping allows us to study different observables like the entanglement entropy. Entanglement entropy is obtained by the mapping $y_{\pm}=F_\pm(x_\pm)$ of arbitrary geodesic in Poincare coordinates with the spacelike separated endpoints $(t_1,y_1)=(y_1^-,y_1^+)$ and $(t_2,y_2)=(y_2^-,y_2^+)$. In the Poincare patch the length of the geodesic between these points is expressed as
\be 
S=\frac{c}{12}\log\left(\frac{(y_2^+-y_1^+)^2(y_2^--y_1^-)^2}{u_{1\delta}u_{2\delta}}\right),
\ee 
where $u_{1\delta}$  and $u_{2\delta}$ are the divergences corresponding to the each of the geodesic endpoints.
In principle it is enough to know the near-boundary mapping $F$ between the Poincare patch and the Banados geometry to get  the entanglement entropy for a fixed region. For the excited state defined by  particular $F$ we have
\be \label{banS}
S=\frac{c}{12}\log\left(\frac{(F(x_2^+)-F(x_1^+))^2(F(x_2^-)-F(x_1^-))^2}{F'(x_2^+)F'(x_1^+)F'(x_2^-)F'(x_1^-)z_{1\delta}z_{2\delta}}\right),
\ee 
where $z_{1\delta}, z_{2\delta}$ are geodesic cut-offs.  Now it is straightforward to obtain the entanglement contour for the interval $(-\ell/2,\ell/2)$ at the time moment  $t$ using formula \eqref{partS} 
\begin{gather}
f(\ell,t)=\frac{c}{6}  \Big(\frac{\left(F\left(\frac{\ell
   }{2}-t\right)-F\left(-t-\frac{\ell }{2}\right)\right)
   F'(x-t)}{\left(F(x-t)-F\left(-t-\frac{\ell }{2}\right)\right)
   \left(F\left(\frac{\ell
   }{2}-t\right)-F(x-t)\right)}+\frac{\left(F\left(t+\frac{\ell
   }{2}\right)-F\left(t-\frac{\ell }{2}\right)\right)
   F'(t+x)}{\left(F(t+x)-F\left(t-\frac{\ell }{2}\right)\right)
   \left(F\left(t+\frac{\ell }{2}\right)-F(t+x)\right)}\Big).
\end{gather}
This entanglement contour is completely specified by $F$ and its derivatives.  For the choice
\be 
F_\pm=\exp(x_\pm/z_h),
\ee 
 we reproduce the entanglement contour for the single interval at finite temperature 2d CFT. If the state is time-independent (however, admitting inhomogeneous  distribution of the energy), the expression for the contour simplifies to 
 \be 
 f(\ell)=\frac{c}{3}\frac{
   \left(F\left(\ell/2\right)-F\left(-\ell/2\right)\right) }{
   \left(F\left(\ell/2\right)-F(x)\right)
   \left(F(x)-F\left(-\ell/2\right)\right)}F'(x).
 \ee
 The Banados geometry can be used to describe many non-trivial states in 2d CFT including global quenches, local quenches, states after partial projective measurements and the others. As a particular example  consider the state corresponding to the holographic boundary CFT at finite volume and temperature (at large central charge) \cite{Mandal:2016cdw}. This case corresponds  to the mapping by certain elliptic functions that compactifies the upper-half plane to the finite size rectangle. We take this mapping to be
\begin{gather} \label{finite}
\frac{\beta}{L}=\frac{4K(b^4)}{K(1-b^4)},\,\,\,\,\,\,
F(x)=-b\frac{ \text{sn}\left(\frac{4 
   K\left(b^4\right)}{\beta }\left(x-L/2\right)|1-b^4\right)}{\text{cn}\left(\frac{4
    K\left(b^4\right)}{\beta
   }\left(x-L/2\right)|1-b^4\right)},
\end{gather}
following the analysis from \cite{Mandal:2016cdw} (see also \cite{Kuns:2014zka,deBoer:2016bov}).
  Typically there are more than one  geodesic configurations competing with \eqref{banS}. These configurations correspond to different conformal block channels expansion of the entanglement entropy. We plot the typical entanglement evolution in Fig.\ref{fig:ds} and  the evolution of the entanglement contour with all geodesic contributions taken in account in Fig.\ref{fig:osc1}. The picture of entanglement evolution is the following: we see the oscillating behavior that is suddenly violated by dips and ramps  corresponding to times where the dominating geodesic configuration changes.   In Fig.\ref{fig:osc1}, we draw the difference between the entanglement contour of excited state \eqref{finite} and the CFT entanglement contour at zero temperature with the same fixed $L$. We see how, from the center of the interval
  propagates the kind of entanglement tsunami which reflects from the interval boundaries after some time. This shows how revivals that are typical to the finite volume and temperature are distributed inside the interval and corresponding degrees of freedom. In Fig.\ref{fig:osc2}, we plot the contribution only from one channel and compare it with other proposals defined below in the text.

\subsection{Thermodynamics of  entanglement contour}\label{entE2}
It is known that the entropy counts the number of
microstates and  this is a fundamental law that relates the information contained in a system and its total
energy. So there is a natural question of how the entanglement entropy is related to the thermodynamic quantities of the system.
 In the context of the holographic approach, the entanglement being expressed by HRRT formula through the geometric measure, such as the minimal surfaces in a certain background, gives us insight into this relation, as well as a new understanding of gravity nature. The entanglement entropy calculated in the Banados spacetime is fixed by  
 \be \label{sat}
 S''+\frac{6}{c}(S')^2=2\pi T_{++},
 \ee 
 where $T_{++}$ is the null component of the stress-energy tensor in the dual system\footnote{Here prime denotes the derivative with respect to null coordinates}. This equation corresponds to  the quantum null energy condition  saturation  for Banados geometry \cite{Ecker:2019ocp}. It is interesting to consider how the entanglement contour of some quite general excited state is related to the energy distribution in the system. Following \cite{Bhattacharya:2013bna,Nozaki:2013vta}  let us consider the perturbation
 \be 
 g_{\mu\nu}=\eta_{\mu\nu}+h_{\mu\nu},
 \ee 
 of the $AdS_3$ in the Fefferman-Graham gauge
 \be 
 ds^2=\frac{dz^2+g_{\mu\nu}dx^{\mu}dx^{\nu}}{z^2},
 \ee 
 by the (arbitrary) scalar matter with the Einstein equations to be hold.
 In \cite{Bhattacharya:2013bna,Nozaki:2013vta} it was shown, that in the  small interval ($x\in(x_0,x_1)$) limit the entanglement (with the ground state entanglement being extracted) in this setup is fixed by the  relation
  \be \label{S-T}
\Delta S=\frac{L}{24 G}\ell^2 T_{tt}(t,\xi),\,\,\xi=\frac{(x_0+x_1)}{2},\,\,\ell=x_1-x_0.
 \ee 
 Thus it is straightforward to obtain the entanglement contour from \eqref{S-T} for the interval $x\in(-\ell/2,\ell/2)$. This entanglement contour has the form
\begin{gather}
 a=2x+\ell,\,\,\,b=2x-\ell,\\
f(x)\sim a T_{tt}\left(t,\frac{x}{2}-\frac{\ell }{4}\right)--b T_{tt}\left(t,\frac{(2 x+\ell )}{4} \right)+\\\frac{a^2}{8} T_{tt}^{(0,1)}\left(t,\frac{x}{2}-\frac{\ell
   }{4}\right)-\frac{b^2}{8} 
   T_{tt}^{(0,1)}\left(t,\frac{1}{4} (2 x+\ell )\right),
   \end{gather}
where index $(0,1)$ denotes the derivative with respect to the second (spatial) argument.

\section{The geometric contour measures}
\label{GCM}
Formula \eqref{RT} defines the entanglement entropy of the region $A$ as the minimal surface ${\cal E}_A$ spanned on the boundary of this region. It is natural to try to give the definition of the entanglement contour which we call geometric entanglement contour(GEC) as the density of the induced (one-dimensional) metric on ${\cal E}_A$. 
Note that in \cite{Wen:2018whg}, the precise correspondence relating the points on the geodesic and the entanglement contour has been given for particular background. In some sense, what we do is the extension of the simplified version of this correspondence on the non-equilibrium backgrounds. What we find is that the GEC qualitatively resembles contour corresponding to one of the channels from partial entanglement entropy proposals. It would be interesting to extend our computations and use the precise correspondence in the non-equilibrium situation.

GEC definition suffers from the fact that one has to introduce the "dynamical" cutoff depending on the size of the subsystem. Also, it makes sense only for  connected subsystems like single intervals. While all these disadvantages take place, this definition of the contour function is quite natural and geometric, giving insight into the entanglement structure.   We show that it resembles the structure of the entanglement contour for specific regions and states. Our main goal here is to study such a proposal for the non-stationary situation.
In general, GEC   consists of the function $g_A(x)$ playing the same role as the EEC and the dynamical cutoff $\delta_A$\footnote{Here we are going to focus on $g_A(x)$}. 

$\,$

Consider the main examples of $AdS_3/CFT_2$ correspondence starting with the Poincare patch, then consider the BTZ black hole, global $AdS_3$ and the deformation of it dual to excited state created by the insertion of the primary operator.
The simplest solution of three-dimensional gravity with the cosmological constant is the Poincare patch of $AdS_3$ with the metric
\be 
ds^2=\frac{L^2}{z^2}\left(-dt^2+dx^2+dz^2\right).
\ee 
The geodesic spanned on  the endpoints of the interval of  length $\ell$ (for definiteness take $x\in(-\ell/2,\ell/2)$) is given by 
\be 
z_P(x)=\sqrt{\ell^2/4-x^2},
\ee 
and the corresponding GEC $g(x)$ reads as
\begin{gather}
ds=L\frac{\sqrt{1+z^{\prime}(x)^2}}{z(x)}dx=g(x)dx,\\g(x)=\frac{2\ell L}{\ell^2-4x^2},\,\,\, \delta_A=1/\ell,
\end{gather}
which resembles the known answer for the single interval in 2d CFT. It is straightforward to compute the GEC for other basic examples. Leaving  details to  \ref{BasisEC}  we find that the GEC for the thermal state is given by
\be
f^{\text{BTZ}}(x)=\frac{\pi}{\beta}\left(\coth \left(\frac{\pi  (\ell-2 x)}{\beta
   }\right)+\coth \left(\frac{\pi  (\ell+2 x)}{\beta
   }\right)\right).
\ee 
It is worth to notice, that GEC (as well as entanglement contour given by partial entanglement entropy) resembles the kernels of modular Hamiltonian for corresponding states. In \cite{Wen:2018whg,Wen:2019ubu} it was shown,  how to reconstruct the reduced density matrix using entanglement contour proposal. Moreover, it was shown how to determine local modular flow  and how it relates points on the boundary interval and on the RT surface. This makes the similarity of the GEC and the related kernels quite natural.

 In this section, we propose how to define GEC for Banados geometries using only the mapping from the Poincare $AdS_3$ and reproduce the answer for the thermal system as a particular example. 
As it is easy to see from the previous subsection the induced metric on the RT surface naturally defines the GEC. Using $F_\pm$ the RT surface in the Poincare coordinates is mapped to the HRT surface in Badanos geometry. HRRT surface is not a necessary constant-time curve. This raises the question of how correspondence works in this general situation. However, let us proceed and see what we get.

First, consider the induced metric on the arbitrary geodesic in the Poincare coordinates with the endpoints at different times $t_1$ and $t_2$. The geodesic connecting points  $(y_1,t_1)$ and $(y_2,t_2)$ is given\footnote{Here we assume $y_2-y_1=\Delta y>0$ and $\Delta t=t_2-t_1<\ell$, with $\Delta t>0$}  by
\begin{gather}
u(y)=\frac{\sqrt{\left(y_1-y\right) \left(y-y_2\right) \left(\Delta t+\Delta y\right)
   \left(\Delta y-\Delta t\right)}}{\Delta y},\\
t(y)= \frac{\left(t_1-t_2\right) }{y_1-y_2}y+\frac{t_2 y_1-t_1 y_2}{y_1-y_2}.
\end{gather}
The induced metric on this geodesic is given 
by 
\be 
ds=\frac{1}{2} \left(\frac{1}{y-y_1}+\frac{1}{y_2-y}\right)dy,
\ee 
which is independent of $t_1$ and $t_2$ and coincides with the  induced metric on the constant-time geodesic and the corresponding GEC. Rewriting it in the lightcone coordinates $y_\pm$ we obtain
\begin{gather}
ds=\frac{1}{2}(dy_++dy_-)\Big(\frac{1}{y_-+y_+-y_{-1}-y_{+1}}+\frac{1}{-y_--y_++y_{-2}+y_{+2}}\Big).
\end{gather}
After choice of the mapping $y_\pm=F_\pm(x_\pm)$ we get that
\begin{gather}
ds=\frac{1}{2}\Big(\frac{(F'(x_+)dx_+ + F'(x_-)dx_-)}{F_-(x_-)+F_+(x_+)-F_-(x_{-1})-F_+(x_{+1})}+\frac{(F'(x_+)dx_+ + F'(x_-)dx_-)}{-F_-(x_-)-F_+(x_+)+F_-(x_{-2})+F_+(x_{+2})}\Big).
\end{gather}
Now we return to  $x$ and $t$ coordinates and set $t_1=t_2=t$. This means that we consider  the geodesic endpoints in the same constant time slice in Banados metric. After that we take constant time slice section  $dt=0$ and  obtain the general expression for the GEC in the form 
\begin{gather}
g(x)\sim\frac{1}{4}
   \Big(\frac{2(F'(x-t)+F'(t+x))}{F\left(\frac{\ell}{2}-t\right)+F\left(\frac   {\ell}{2}+t\right)-F(x-t)-F(t+x)}-\\ \frac{2(F'(x-t)+F'(t+x))}{F\left(-\frac
   {\ell}{2}-t\right)+F\left(t-\frac{\ell}{2}\right)-F
   (x-t)-F(t+x)}\Big).
\end{gather} 
This formula reproduces GEC for thermal CFT if we choose the mapping corresponding to the BTZ black hole $F(x)=\exp(2/z_h)$. In Fig.\ref{fig:osc1} we plot  the GEC corresponding to the mapping \eqref{finite}. We see that  the evolution of the system is qualitatively the same as for the entanglement contour.

\section{The remarks on the complexity contour and geometric coarse-graining}
\label{CM}
The contour measures of entanglement give us insight to the entanglement(or another measure)  distributed inside the subsystem. It is hard to fix "the best" contour measure. The most basic and straightforward requirements are constraints by the normalization and positivity. There are different ways how to introduce the contour function in holography. It seems that the most straightforward is the partial entanglement proposal for the contour \cite{Wen:2018whg,Wen:2019ubu,Kudler-Flam:2019oru}. The application of this proposal makes the non-equilibrium picture of entanglement evolution more coarse-grained as, for example, in \cite{Kudler-Flam:2019oru}. However, the HRRT formula is  geometric, and it is likely for the contour function also to have the geometric nature.  This can be made more precise being tested on another quantum-informational measure holographic calculation of which is related to the entanglement and HRRT surfaces. As an example we give  the well-known CV complexity where the non-geometric proposal fails to describe any coarse-graining for lower-dimensional systems.
On the other hand, it seems that the usage of the geometrical proposals like integrated volume defined by the HRRT or bit-thread picture can give us a more precise picture. We leave this as a comment, and we are not going to discuss this in full generality. A detailed investigation of these questions will be given somewhere else.

$\,$

The "complexity' is a relatively new measure in the holographic quantum-information realm. Roughly speaking, this measure describes quantitatively the answer to the question, "how complicated is the process of obtaining some state in the quantum system? Taking into account the importance of the entanglement contour, it is natural to define the notion of {\it "complexity contour"} for a given subsystem.

Define the "volume complexity"  $\cC_{\cV}$ using codimension-one bulk hypersurface $B$ attached to the fixed time slice of the boundary
 \bea
\cC_{\cV}(\Sigma)=\frac{\cV(B)}{G L},
 \eea
where $\cV(B)$ is the volume of this hypersurface, and $G$ is the gravitational constant. 
The CA duality relates the complexity of the state to the value of the on-shell gravitational action restricted to the Wheeler-DeWitt (WDW) patch
 \bea
\cC_{A}=\frac{S({\cal W})}{\pi }.
 \eea
 WDW patch $\cal W$ is defined as the bulk domain of dependence of any Cauchy surface asymptotically approaching the fixed time slice of the boundary.   As in the previous sections, we focus on the complexity of the subregion chosen to be the single interval.
  The conventional formulation (especially the covariant one) of CV or CA complexity conjectures for subregions is absent at the moment. We will focus on the following versions of these prescriptions. The CV conjecture for subregions that we consider relates the volume restricted by the RT surface associated with some subregion $A$.  This proposal equates the complexity and on-shell gravitational action evaluated in the intersection of the entanglement wedge and the WdW patch.  The CA complexity is characterized by complicated divergences structure, which has to be taken into account properly.
  
  In this paper, we focus on the CV complexity to avoid additional considerations related to additional divergences. The volume complexity of the interval with the length $\ell$ is independent of temperature\footnote{However there is non-trivial dependence due to topological effects, see \cite{Abt:2017pmf}}, and the dependence on the interval length is given in the form
 \be 
{ \cC}\sim\ell/\varepsilon,
 \ee   
where $\varepsilon$ is the divergence. In contrast to this fact, the entanglement exhibits explicit temperature dependence. If one will try to proceed by analogy with the construction similar to the partial entanglement entropy, the result will also be temperature independent. The dependence of complexity contour constructed in this way looks unsatisfactory (however, this does not exclude the possibility that this is the "right" contour).

To get the insight into how the complexity degrees of freedom are distributed inside the interval, it is natural to proceed in the spirit of GEC and define contour-like quantity.  For example, the natural geometric contour measure $f_V(x)$, in this case, could be the integrated volume at the point $x$. Considering the BTZ black hole and the volume enclosed by the geodesic $z_{BTZ}$ (see \eqref{zbtz} for the explicit form) we obtain
\be 
f_V(x)=\int_\varepsilon^{z=z_{BTZ}(x)}\frac{dz}{z^2\sqrt{f(z)}},
\ee 
which is temperature-dependent. Of course, this way of definition of the contour functions could also suffer from the problem that we have to take care of divergences coming from integration over $x$, making an additional change of variables as in the GEC. This definition strongly depends on the entanglement surface, thus providing that the complexity has the non-trivial distribution inside the interval. To make the connection to the more well-established contour proposal, one can try to perform the integration over the integral curves of bit-thread flows, which are known are to define the entanglement contour \cite{Kudler-Flam:2019oru}.  It is worth noting that bit-thread proposal mapping coincides with Wen's fine-grained correspondence in simple cases. Also, it would be interesting to make the computation of the evolution of complexity contours in a non-equilibrium situation like local quenches \cite{Ageev:2019fxn}.

\section*{Acknowledgements}
Author would like to thank Ian MacCormack, Jonah Kudler-Flam and Qiang Wen  for  correspondence  and discussion.  Also I would like to thank Yu. Ageeva for careful reading of the early version of this manuscript and I.Ya. Aref'eva for support. This work was funded by Russian Federation represented by the Ministry of Science and Higher Education (grant number 075-15-2020-788).

\newpage

\appendix

\section{ The details on basic examples of  geometric entanglement contours}\label{BasisEC}
 In this appendix we find different basic examples of the geometric entanglement contours for the states dual to locally $AdS_3$ spaces. In general the geometric entanglement contour is defined by the induced metric on the geodesic.
 
 First  consider entanglement contour  for the  state dual to the one-sided BTZ black hole
\bea \label{zbtz}\nn
&& ds^2=\frac{L^2}{z^2}\left(-f(z)dt^2+\frac{dz^2}{f(z)}+dx^2\right),\\ \nn
&&z_{\text{BTZ}}(x)=\frac{z_h}{\sqrt{2}} \text{sech}\left(\frac{\ell}{2 z_h}\right)
   \sqrt{\cosh \left(\frac{\ell}{z_h}\right)-\cosh \left(\frac{2
   x}{z_h}\right)},
\eea 
where $f(z)=1-z^2/z_h^2$. We find that GEC has the form
\be 
f^{\text{BTZ}}(x)=\frac{\pi}{\beta}\left(\coth \left(\frac{\pi  (\ell-2 x)}{\beta
   }\right)+\coth \left(\frac{\pi  (\ell+2 x)}{\beta
   }\right)\right).
\ee
The metric dual to the state created by the primary operator insertion (on the circle) is
\be 
ds^2=-\left(A^2+\frac{r^2}{L^2}\right)dt^2+\frac{dr^2}{A^2+\frac{r^2}{L^2}}+r^2d\phi^2,\,\,\,\,\,\,|A|<1,
\ee 
which coincides with the global $AdS_3$ for $A=1$ and it is the space with conical defect for $|A|<1$.
 The geodesic in this space is given by
\be 
r(\phi)=\frac{A L \sec (A \phi )}{\sqrt{\tan ^2\left(\frac{A \ell}{2}\right)-\tan
   ^2(A \phi )}},
\ee 
defining the GEC  in the form
\be
f(\phi)=\frac{A L \sin (A \ell)}{\cos (2 A \phi )-\cos (A \ell)},\,\,\,\,\, A=\sqrt{1-\frac{24 \Delta  L^2}{c}},
\ee 
where $\Delta$ is the weight of the perturbing operator and $c$ is the central charge.

%\begin{acknowledgements}
%If you'd like to thank anyone, place your comments here
%and remove the percent signs.
%\end{acknowledgements}

% BibTeX users please use one of
%\bibliographystyle{spbasic}      % basic style, author-year citations
%\bibliographystyle{spmpsci}      % mathematics and physical sciences
%\bibliographystyle{spphys}       % APS-like style for physics
%\bibliography{}   % name your BibTeX data base

% Non-BibTeX users please use

\end{document}